# A Four Component Decomposition of the Instantaneous Velocity in Turbulent Flow

**Trinh, Khanh Tuoc**

**Institute of Food Nutrition and Human Health**

**Massey University, New Zealand**

*K.T.Trinh@massey.ac.nz*

# Abstract

A four component decomposition of the local instantaneous velocity is proposed. It brings out more readily the terms in the Navier-Stokes equations associated with different events and fluid structures of turbulent flows than the classic two component decomposition of Reynolds. In particular the new composition highlights the existence of two types of Reynolds stresses: fast and slow. The fast Reynolds stresses can be linked to a streaming process that describes the ejection of wall fluid in the bursting process. It also provides a simple method for modelling the wall layer. The four components are a long time average, a slow fluctuating component based on the difference between the long term average and the smoothed phase velocity developed by passing coherent structures, a fast fluctuating component which is periodic in nature and a streaming component created by the interaction between the fast fluctuations and the fluid viscous effects.

# Introduction

Turbulence is a complex time dependent three-dimensional motion widely believed to

be governed by equations[1] established independently by Navier and Stokes more than 150 years ago

$$\frac{\partial}{\partial t}(\rho u_i) = -\frac{\partial}{\partial x_i}(\rho u_i u_j) - \frac{\partial}{\partial x_i}p - \frac{\partial}{\partial x_i}\tau_{ij} + \rho g_i \qquad (1)$$

and the equation of continuity

$$\frac{\partial}{\partial x_i}(\rho u_i) = 0 \qquad (2)$$

This fascinating problem has occupied some of the best scientific minds of the last century and a half but a formal solution is yet to be published.

The omnipresence of turbulence in many areas of interest such as aerodynamics, meteorology and process engineering, to name only a few, has nonetheless led to a voluminous literature based on semi-theoretical and empirical solutions and investigations of selected aspects of turbulence structure and mechanisms. According to the Web of Science electronic database, over 3500 papers were published last year alone. It is a challenge to simply keep abreast of the information!

Most of the interest in turbulence modelling from a practical engineering view point was originally based on the time averaged parameters of the steady state flow field. Reynolds (1895) has proposed that the instantaneous velocity $u_i$ at any point may be decomposed into a long-time average value $U_i$ and a fluctuating term $U'_i$.

$$u_i = U_i + U'_i \qquad (3)$$

with

$$U_i = \lim_{t \to \infty} \int_0^t u_i \, dt \qquad (3.1)$$

$$\int_0^\infty U'_i \, dt = 0 \qquad (3.2)$$

---

[1] The suffices i and j in this paper refer to standard vector notation.

For simplicity, we will consider the case when

    1. The pressure gradient and the body forces can be neglected

    2. The fluid is incompressible ($\rho$ is constant).

Substituting equation (3) into (1) and taking account of the continuity equation (2) gives:

$$U_i \frac{\partial U_j}{\partial x_j} = \nu \frac{\partial^2 U_i}{\partial x_i} - \frac{\partial \overline{U'_i U'_j}}{\partial x_i} \qquad (4)$$

These are the famous Reynolds equations (Schlichting, 1960, p. 529) also called Reynolds-Averaged-Navier-Stokes equations RANS (Gatski & Rumsey, 2002; Hanjalić & Jakirlić, 2002). The long-time-averaged products $\overline{U'_i U'_j}$ arise from the non-linearity of the Navier-Stokes equations. They have the dimensions of stress and are known as the Reynolds stresses. They are absent in steady laminar flow and form the distinguishing features of turbulence.

The writer proposes that the traditional picture implied by the RANS is an oversimplification and that more information about the Reynolds stresses can be obtained by a more detailed analysis.

## A more detailed decomposition of the instantaneous velocity

The derivation of equation (4) implies a velocity trace with a stationary long-time average as shown in Figure 1. Reynolds further imagined the fluctuating components $U'_i$ to be random.

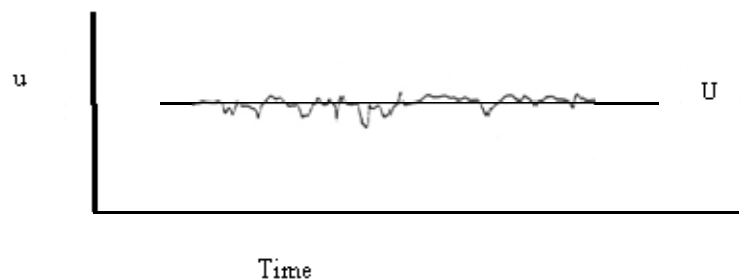

Figure 1. Decomposition of the streamwise component of the instantaneous velocity

according to Reynolds (1895), Data of Antonia et al.(1990).

The advance in measuring techniques of the last sixty years have shown conclusively that the instantaneous velocity traces of flow close to a wall show two types of fluctuations: fast and slow. Figure 2 shows a typical trace of streamwise velocity near the wall, redrawn after the measurements of Antonia, Bisset, & Browne (1990). If we draw a smooth line through this velocity trace so that there are no secondary peaks within the typical timescale of the flow $t_v$, we define a locus of smoothed velocity $\tilde{u}_i$ and fast fluctuations $u'_i$ of period $t_f$ relative this base line.
Then we may write

$$\int_0^\infty u'_i dt = 0 \qquad (5)$$

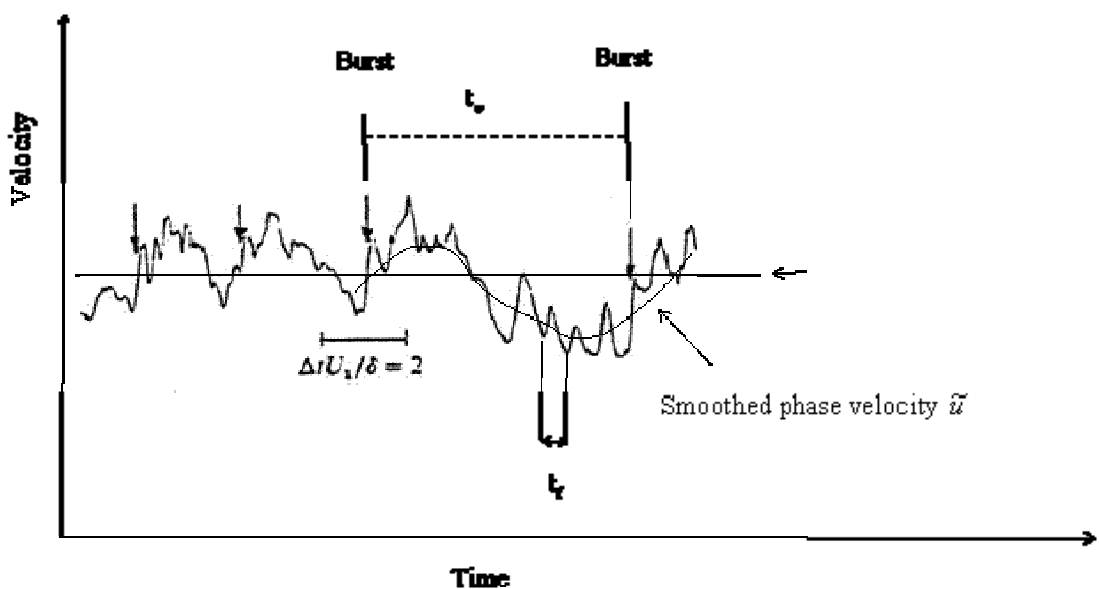

Figure 2. Trace of instantaneous streamwise velocity after measurements by Antonia et al. (1990).

H. T. Kim, Kline, & Reynolds (1971), for example, have obtained the distribution of the smoothed instantaneous streamwise velocities near the wall by conditional sampling at various phases of the bursting cycle (Figure 3)

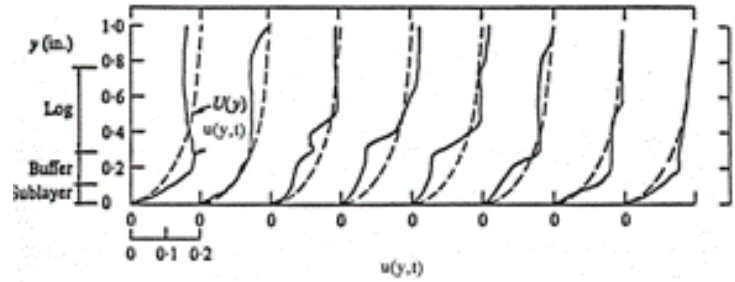

. Figure 3 Smoothed phase velocity in a bursting cycle according to Kim et al. (1971)

Mankbadi (1992) also defines the conditional average in the same way as the phase average:

$$\tilde{u}(y,t) = \lim_{N \to \infty} N^{-1} \sum_{n=0}^{N} u(y, t + nt_v) \qquad (6)$$

It is of course necessary to detect first the beginning of an event and determine its characteristic time scale $t_v$. Thus Antonia (1980) defines the conditionally averaged velocity as:

$$\tilde{u}_i = \lim_{t_v \to \infty} \left[ \frac{\int_0^t u_i c(t)\, dt}{\int_0^t c(t)\, dt} \right] \qquad (7)$$

It is a measure of the smoothed instantaneous velocity at a particular phase of an event. To avoid confusion in the nomenclature, we will call this "the smoothed phase velocity". Antonia (1980) discusses various detection schemes used to define the function $c(t)$ that presumably locks the sampling onto a special feature associated with the coherent structure.

We draw two conclusions from the work of Kim et al:
1. The fast fluctuations are eliminated by the conditional sampling process.
2. The long-time-averaged velocity profile monitored by the Reynolds

equations does not correspond to the smoothed phase velocity at any instant in time. Thus some information is lost in the method of velocity decomposition proposed by Reynolds.

The decomposition of the velocity into fast and slow fluctuations brings out more readily the transient structures of the flow and is crucial to the success of the large eddy simulations LES, direct numerical simulations DNS and the variable-interval time-averaging technique, VITA, of Blackwelder & Kaplan (1976).

The instantaneous velocity in the sweep phase of the wall process may be decomposed in an alternate manner as:

$$u_i = \tilde{u}_i + u'_i \tag{8}$$

Comparing equations (2) and (8) shows that

$$U_i = \frac{1}{t_\nu} \int_0^{t_\nu} \tilde{u}_i \, dt \tag{9}$$

and

$$U'_i = \tilde{U}'_i + u'_i \tag{10}$$

where

$$\tilde{U}'_i = \tilde{u}_i - U_i \tag{11}$$

then

$$u_i = U_i + \tilde{U}'_i + u'_i \tag{12}$$

## Reynolds stresses

We may average the Navier-Stokes equations over the period $t_f$ of the fast fluctuations. Bird, Stewart, & Lightfoot (1960), p. 158 give the results as

$$\frac{\partial (\rho \tilde{u}_i)}{\partial t} = -\frac{\partial p}{\partial x_i} + \mu \frac{\partial^2 \tilde{u}_i}{\partial x_j^2} - \frac{\partial \tilde{u}_i \tilde{u}_j}{\partial x_j} - \frac{\partial \overline{u'_i u'_j}}{\partial x_j} \tag{13}$$

Equation (13) defines a second set of Reynolds stresses $\overline{u'_i u'_j}$ which we will call "fast" Reynolds stresses to differentiate them from the standard Reynolds stresses $\overline{U'_i U'_j}$. In general $u'_i < U'_i$ and the fast Reynolds stresses are smaller in magnitude than the standard Reynolds stresses.

To the writer's knowledge experimental investigations of turbulence, up to the time he first presented this theory to colleagues in Australasia (Trinh, 1992), all targeted the standard Reynolds stresses and no separate measurements existed for the fast Reynolds stresses. Considerations of this second set of Reynolds stresses gives a much better overall picture of the problem, in particular of the causal relationships in the study of the flow structure.

Within a period $t_\nu$, the smoothed velocity $\widetilde{u}_i$ varies slowly with time but the fluctuations $u'_i$ may be assumed to be periodic with a timescale $t_f$. In the particular case of steady laminar flow, $\widetilde{u}_i = U_i$ and $\widetilde{U}'_i = 0$: only the fast fluctuations remain. These are typically remnants of disturbances introduced at the pipe entrance or leading edge of a flat plate by conditions upstream.

We may write the fast fluctuations in the form
$$u'_i = u_{0,i} \left( e^{i\omega t} + e^{-i\omega t} \right) \tag{14}$$

The fast Reynolds stresses $u'_i u'_j$ become
$$u'_i u'_j = u_{0,i} u_{0,j} \left( e^{2i\omega t} + e^{-2i\omega t} \right) + 2 u_{0,i} u_{0,j} \tag{15}$$

Equation (15) shows that the fluctuating periodic motion $u'_i$ generates two components of the "fast" Reynolds stresses: one is oscillating and cancels out upon long-time-averaging, the other, $u_{0,i} u_{0,j}$ is persistent in the sense that it does not depend on the period $t_f$. The term $u_{0,i} u_{0,j}$ indicates the startling possibility that a purely oscillating motion can generate a steady motion which is not aligned in the direction of the oscillations. The qualification steady must be understood as independent of the frequency $\omega$ of the fast fluctuations. If the flow is averaged over a longer time than the period $t_v$ of the bursting process, the term $u_{0,i}u_{0,j}$ must be understood as transient but non-oscillating. This term indicates the presence of transient shear layers embedded in turbulent flow fields and not aligned in the stream wise direction similar to those associated with the streaming flow in oscillating laminar boundary layers (Schneck & Walburn, 1976; Tetlionis, 1981).

**Coherent structures near the wall**

Oblique shear layers have been observed near the wall and upstream of large scale structures by (R.A. Antonia, Browne, & Bisset, 1989; Blackwelder & Kovasznay, 1972; Brown & Thomas, 1977; Chen & Blackwelder, 1978; Falco, 1977; Hedley & Keffer, 1974; Nychas, Hershey, & Brodkey, 1973; Spina & Smits, 1987). These structures are characteristic of patches of fluid that move within turbulent flow fields. An extraordinary number of these structures have been identified in the past five decades prompting one researcher to say "When studying the literature on boundary layers, one is soon lost in a zoo of structures, e.g. horseshoe- and hairpin-eddies, pancake- and surfboard-eddies, typical eddies, vortex rings, mushroom-eddies, arrowhead-eddies, etc…"(Fiedler, 1988) It is not clear from literature reports whether different observations refer exactly to the same phenomenon and what effects the different methods of event detection have on the results.

In fact the first observations of coherent structures date from Reynolds (1883). Interest in these structures was reignited by the classic work of Kline et al. (1967). Using hydrogen bubbles as tracers, they observed inrushes of high-speed fluid from the outer region towards the wall, followed by longitudinal sweeps along the wall. During the sweep phase, the structure of the wall layer shows alternate streaks of high and low-speed fluid. The low-speed streaks become unstable, lift and oscillate until they are eventually ejected into the outer region in a violent burst. Kline et al. observed that the hydrogen bubble lines in their experiments became contorted during the ejection phase indicating a break-up of the flow into small scales. They refer to the wall-layer process at this point as bursting. Most of turbulent stresses in the wall layer are produced during this short bursting phase compared with the much longer sweep phase. The work of Kline et al. highlighted the transient nature of the wall layer process and the existence of a secondary stream when most of the turbulent stresses were produced.

Because of the importance of the wall region as highlighted by the work of Kline et al., a large amount of effort has been devoted to its study focussing mainly on the hairpin vortex, the most identifiable coherent structure in that region. Work before 1990 were well reviewed, for example by Cantwell (1981) and Robinson (1991). There have been many physical experiments e.g. (Blackwelder & Kaplan, 1976; Bogard & Tiederman, 1986; Carlier & Stanislas, 2005; Corino & Brodkey, 1969; Head & Bandhyopadhyay, 1981; Luchak & Tiederman, 1987; Meinhart & Adrian, 1995; Tardu, 1995; A. A. Townsend, 1979; Willmarth & Lu, 1972), including efforts to induce artificially the creation of a hairpin vortex by injecting a jet of low momentum fluid into a laminar flow field (Arcalar & Smith, 1987; Gad-el-Hak & Hussain, 1986; Haidari & Smith, 1994). With the advent of better computing facilities, direct numerical simulations DNS have been used increasingly to conduct 'numerical experiments" e.g. (Jimenez & Pinelli, 1999; J. Kim, Moin, & Moser, 1987; Spalart, 1988).

Much more temporal detail can be deduced from numerical experiments. For example, Johansson, Alfresson, & Kim (1991) analysed the data base provided by the DNS of Kim, Moin and Moser (1987) to obtain the conditionally averaged production of turbulent kinetic energy $\widetilde{P}$ which they write as

$$\widetilde{P} = \overline{U'V'}\frac{dU}{dy} - \overline{U'V'}\left(\frac{\partial \widetilde{U}'}{\partial x} + \frac{\partial \widetilde{V}'}{\partial y}\right) - \overline{V'^2}\frac{\partial \widetilde{V}'}{\partial y} - \overline{W'^2}\frac{\partial \widetilde{W}'}{\partial z} \qquad (16)$$

The first term on the right-hand side of equation (16) is the only one that remains in the long-time averaged sense. It is shown in Figure 4b. The total conditionally averaged production $\widetilde{P}$ is substantially higher as seen in Figure 4a. The difference between these two terms is shown in Figure 4c. It points to the existence of an important transient contribution weakly slanted with respect to the wall and which can be attributed to strong gradients in the x- and y- directions of the conditionally averaged streamwise velocity.

There has been a slow build up of view that the destabilisation of a laminar flow field cannot be simply explain in terms of growth of periodic disturbances alone as originally investigated by many authors e.g. Tollmien (1929), Schubauer & Skramstad (1943), Schlichting (1960) and must involve a second mechanism e.g. Trefethen, Trefethen, Reddy, & Driscoll (1993). Schoppa & Hussain (2002). Schoppa and Hussain have analysed their DNS data base to argue that sinusoidal velocity fluctuations led to the production of intense shear layers associated with the streaming flow, that they call transient stress growth TSG. They attribute the lifting of the longitudinal wall vortex into the head of a hairpin vortex directly to the action of the TSG.

Rather than rely on very detailed and complex arguments based on the analysis of vorticity patterns obtained from DNS, PIV (particle imaging velocimetry) or velocity probe measurements with different detection schemes, the writer prefers to use a technique borrowed from the study of laminar oscillating boundary layers (K.T. Trinh, 1992) to identify the different terms in the Navier-Stokes equations related to different structures and their interaction. This approach is similar to the study of Kolmogorv flows, simple sinusoidal flows that Kolmorov advocated as models for investigations into the transition to turbulence. Meshalkim and Sinai (1961). were the first to take up the suggestion followed by many others e.g. (Balmforth & Young, 2002) .

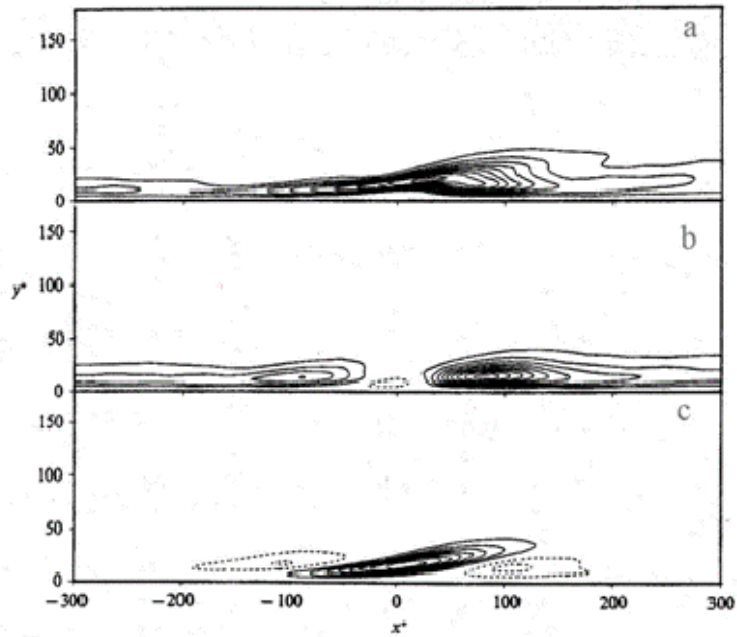

Figure 4. Production of turbulence near the wall. (a) $\tilde{P}$, (b) $\overline{U'V'}(dU/dy)$, (c) $\tilde{P} - \overline{U'V'}(dU/dy)$. After Johansson, Alfresson, & Kim (1991)

**Oscillating Laminar Boundary Layers**

The analysis of oscillating laminar boundary layers also begins with equation (1). The velocity is decomposed into steady and periodic components. These conditions are exactly the same as those adopted in the DNS (J. Kim, et al., 1987; Laurien & Kleiser, 1989; Spalart, 1988) and the writer believes that techniques developed in the former field of research may be transposed to the study of turbulence. The case of oscillating flow with a zero-mean velocity is particularly interesting since the basic velocity fluctuations imposed by external means do not grow with time because there is no mean motion along the wall. One may thus investigate the effect of the amplitude and frequency of the fluctuations separately. The following treatment of the problem is taken from the excellent book of (Tetlionis, 1981).

We define a stream function $\psi$ such that

$$u = \frac{\partial \psi}{\partial y} \qquad v = \frac{\partial \psi}{\partial x} \qquad (17)$$

Where $u, v$ are now the velocity components in the x and y directions. The basic variables are made non-dimensional

$$x^* = \frac{x}{L} \quad y^* = \frac{y}{\sqrt{2\nu/\omega}} \quad t^* = t\omega \tag{18}$$

$$U_e^*(x,t) = \frac{U_e}{U_\infty}(x,t) \quad \psi^* = \psi \left( U_\infty \sqrt{\frac{2\nu}{\omega}} \right)^{-1} \tag{19}$$

where $U_\infty$ is the approach velocity for $x \to \infty$, $U_e$ is the local mainstream velocity and L is a characteristic dimension of the body. The system of coordinates x, y is attached to the body. The Navier-Stokes equation (1) may be transformed as:

$$\frac{\partial^2 \psi^*}{\partial y^* \partial t^*} - \frac{1}{2}\frac{\partial^3 \psi^*}{\partial y^{*3}} - \frac{\partial U_e^*}{\partial t^*} = \frac{U_e}{L\omega}\left( -\frac{\partial \psi^*}{\partial y^*}\frac{\partial^2 \psi^*}{\partial y^* \partial x^*} + \frac{\partial \psi^*}{\partial x^*}\frac{\partial^2 \psi^*}{\partial y^{*2}} + U_e^* \frac{\partial U_e^*}{\partial x^*} \right) \tag{20}$$

with boundary conditions

$$\psi^{*'} = \frac{\partial \psi^*}{\partial y^*} = 0 \quad y^* = 0 \tag{21}$$

For large frequencies, the RHS of equation (20) can be neglected since

$$\varepsilon = \frac{U_e}{L\omega} \ll 1 \tag{22}$$

In this case, Tetlionis reports the solution of equation (20) as:

$$\psi^* = \left[ \frac{U_0^*(x^*)}{2}(1-i)[1-e^{(1+i)y^*}] + \frac{U_0^* y^*}{2} \right] e^{it^*} + C \tag{23}$$

Tetlionis (op. cit. p. 157) points out that equation (23) may be regarded as a generalisation of Stokes' solution (1851) for an oscillating flat plate. This latter solution describes an oscillating flow called the Stokes layer which is often found embedded in other flow fields and has properties almost independent of the host field. Since Stokes also produced a solution for a flat plate started impulsively, often referred to as Stokes' first problem, the oscillating plate will be referred to as the

Stokes solution2 for clarity. Van Driest (1956) has used the Stokes solution2 to model the damping function in Prandtl' mixing-length theory (1935) near the wall.

Equation (23) is accurate only to an error of order $\varepsilon$. Tetlionis reports a more accurate solution for the case when $\varepsilon$ cannot be neglected (i.e. for lower frequencies):

$$\psi^* = \frac{U_0^*(x^*)}{2}[\psi_0^*(y^*)e^{it^*} \overline{\psi_0^*(y^*)}e^{-it^*}] + \varepsilon[\psi_1^*(x^*,y^*)e^{2it^*} + \overline{\psi_0^*}e^{-2it^*}] + O(\varepsilon^2) \qquad (24)$$

where $\psi_0$ and $\psi_1$ are the components of the stream function of order $\varepsilon^0$ and $\varepsilon$. Substituting this more accurate solution into equation (20), we find that the multiplication of coefficients of $e^{it^*}$ and $e^{-it^*}$ forms terms that are independent of the oscillating frequency, ω, imposed on the flow field and were not anticipated in equation (24). Thus the full solution of equation (20) is normally written (Stuart, 1966; Tetlionis, 1981) as

$$\psi^* = \frac{U_0^*(x^*)}{2}[\psi_0^*(y^*)e^{it^*} + \overline{\psi_0^*}(y^*)e^{-it^*}]$$
$$+ \varepsilon[\psi_{st}^* + [\psi_1^*(x^*,y^*)e^{2it^*} + \overline{\psi_1^*}(x^*,y^*)e^{-2it^*}] + O(\varepsilon^2) \qquad (25)$$

where the overbar denotes the complex conjugate and $\psi_{st}^*$ results from cancelling of $e^{it^*}$ and $e^{-it^*}$ terms.

The quantity $\psi_{st}^*$ shows that the interaction of convected inertial effects of forced oscillations with viscous effects near a wall results in a non-oscillating motion that is referred to in the literature as "Streaming". The problem has been known for over a century (Andrade, 1931; Carriere, 1929; Dvorak, 1874; Faraday, 1831; Rayleigh, 1880, 1884; Schlichting, 1932) and studied theoretically (Riley, 1975; Schlichting, 1960; Stuart, 1966; Tetlionis, 1981). The existence of this streaming flow, even in this absence of any mainstream flow, is clearly demonstrated in Figure 5.

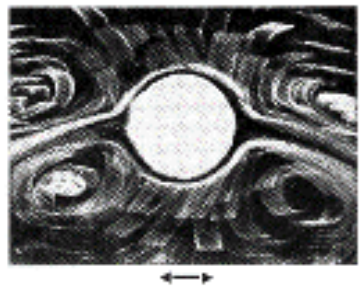

Figure 5. Streaming flow near a vibrating cylinder. After Schlichting (1960).

The governing equation for the streaming function may be extracted from the original Navier-Stokes equations and analysed separately. This is achieved by substituting equation (25) into (20) and collecting the steady terms of order ε. Tetlionis gives the result as

$$-\frac{\partial^3 \psi_{st}^*}{\partial y^{*3}} = -U_0^* \frac{dU_0^*}{dx^*} \frac{\partial \psi_0^*}{\partial y^*} \frac{\partial \overline{\psi_0^*}}{\partial y^*} + \frac{1}{2} U_0^* \frac{dU_0^*}{dx^*} \left( \psi_0^* \frac{\partial^2 \overline{\psi_0^*}}{\partial y^{*2}} + \overline{\psi_0^*} \frac{\partial^2 \psi_0^*}{\partial y^{*2}} \right) + U_0^* \frac{dU_0^*}{dx^*} \quad (26)$$

The boundary conditions imposed in early analyses were:

$$\psi_1^* = \psi_{st}^* = 0 \quad at \quad y^* = 0 \quad (27)$$

$$\frac{\partial \psi_1^*}{\partial y^*} = \frac{\partial \psi_{st}^*}{\partial y^*} = 0 \quad at \quad y^* = 0 \quad (28)$$

$$\frac{\partial \psi_1^*}{\partial y^*} = \frac{\partial \psi_{st}^*}{\partial y^*} = 0 \quad at \quad y^* \to \infty \quad (29)$$

Similarly the governing equation for $\psi_1$ is obtained by collecting the oscillating terms of order ε. The terms of order $\varepsilon^0$ give

$$\frac{\partial^3 \psi_0^*}{\partial y^{*3}} - 2i \frac{\partial \psi_0^*}{\partial y^*} = -i \quad (30)$$

The solution for the main oscillating component $\psi_0$ is the same as equation (23) and may be arranged as

$$\psi_0^* = -\frac{1}{2}(1-i)\left[1 - e^{-(1+i)y^*}\right] + y^* \quad (31)$$

Stuart (1966) has noted that the complementary function of equation (31) is $\left(A + By^* + Cy^{*2}\right)$ where A, B and C are functions of $y^*$. In order to satisfy the boundary condition in equation (29), it is necessary to put both B and C equal to zero.

But then the boundary conditions at the wall cannot be satisfied. Stuart proposes that this anomaly can be remedied by assuming that the derivative $\partial \psi_{st}^*/\partial y^*$ does not reach zero at the outer edge of the Stokes layer but remains finite. Then, assuming C = 0, we obtain

$$\psi_{st}^* = U_0^* \frac{dU_0^*}{dx^*}\left(\frac{13}{8} - \frac{3}{4}y^* - \frac{1}{8}e^{-2y^*} - \frac{3}{2}e^{-y^*}\cos y^* - e^{-y^*}\sin y^* - \frac{1}{2}y^* e^{-y^*}\sin y^*\right) \quad (32)$$

This means that there exist two boundary layers: an oscillating Stokes layer $\delta_s$ and a second layer $\delta_{st}$ created by the intrusion of the streaming flow into the outer inviscid region. Tetlionis estimates the order of magnitude of these two layers as

$$\delta_{st} \approx \frac{L\sqrt{\omega \nu}}{U_\infty} \quad (33)$$

and

$$\delta_s \approx \sqrt{\frac{\nu}{\omega}} \quad (34)$$

$$\frac{\delta_{st}}{\delta_s} \approx \frac{L\omega}{U_\infty} \approx \frac{1}{\varepsilon} \quad (35)$$

Since $\varepsilon$ is small, the streaming layer $\delta_{st}$ is much thicker than the Stokes layer $\delta_s$ as shown in Figure 6.

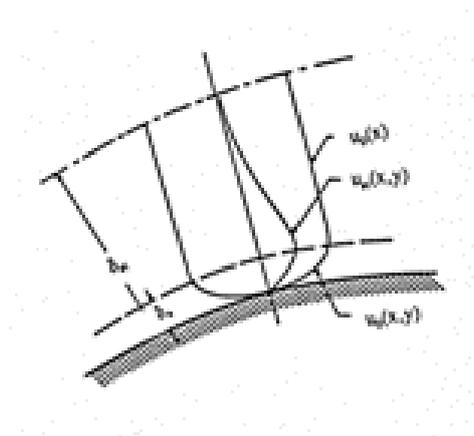

Figure 6. Stokes and Streaming layers. After Tetlionis (1981).

The most important observation is that the streaming flow reaches well beyond the Stokes layer i.e. into the inviscid outer region. This eruption of an unsteady laminar boundary layer is called by various names in kernel studies that attempt to model the wall process of turbulent flow. Peridier, Smith, & Walker (1991) call it viscous-inviscid interaction. These kernel studies arise from the observation that a vortex moving above a wall will induce a laminar sub-boundary layer underneath its path by viscous diffusion of momentum, even if the vortex is introduced into a fluid which was originally at rest (C. R. Smith, Walker, Haidari, & Sobrun, 1991). The vortex impresses a periodic disturbance onto the laminar sub-boundary layer underneath. The problem is thus very similar to that discussed by Tetlionis. In these kernel studies the configuration of the vortex must be specified a priori. In the work of Walker (1978) it is a rectilinear vortex, in Chu and Falco (1988) ring vortices, in Liu et al. (1991) hairpin vortices, in Swearingen and Blackwelder (1987), streamwise Goertler vortices. But recently in their numerical simulation Suponitsky, Cohen, & Bar-Yoseph (2005) have shown that vortical disturbances evolve into a hairpin vortex independently of their original geometry over a wide range of orientations.

The investigation of the flow field outside the Stokes layer has been performed by Stuart (1966) and Riley (1967) using asymptotic expansions. The analyses of Stuart and Riley have the advantage that no assumption need be made about the source of the velocity fluctuations. To order $\varepsilon^0$, the flow in this outer layer is inviscid but the interactions of higher orders are not. The problem is very complex and both workers have introduced an essential simplification: they assume that the streaming flow and the potential flow do not interact. In order to express this simplification mathematically, Stuart has rewritten the stream function in the form

$$\psi = \psi_0(x,t) + y U_e(x,t) + \psi_a(x,y,t) \qquad (36)$$

where $(\psi_0 + y U_e)$ is the periodic potential flow (including a displacement effect) and $\psi_a$ is an additional flow of which we are especially interested in the steady part. The boundary-layer theory is assumed to be valid and the potential flow balances the given pressure gradient. Then equation (20) becomes

$$U_e \frac{\partial^2 \psi_a}{\partial x \partial y} + \frac{\partial U_e}{\partial x} \frac{\partial \psi_a}{\partial y} - \left( \frac{\partial \psi_0}{\partial x} + \frac{y \partial U_e}{\partial x} \right) + \frac{\partial^2 \psi_a}{\partial t \partial y}$$
$$+ \frac{\partial \psi_a}{\partial y} \frac{\partial^2 \psi_a}{\partial x \partial y} - \frac{\partial \psi_a}{\partial x} \frac{\partial^2 \psi_a}{\partial y^2} = \nu \frac{\partial^3 \psi_a}{\partial y^3} \qquad (37)$$

Equation (37) is then averaged with respect to time. The average of $\psi_a$ is denoted by $\psi_{st}^*$

$$\psi_a = \psi_{st} + \psi_t \qquad (38)$$

where $\psi_t$ is the time-dependent part of $\psi_a$. Then we have

$$J + \frac{\partial \psi_{st}}{\partial y} \frac{\partial^2 \psi_{st}}{\partial x \partial y} - \frac{\partial \psi_{st}}{\partial x} \frac{\partial^2 \psi_{st}}{\partial y^2} = \nu \frac{\partial^3 \psi_{st}}{\partial y^3} \qquad (39)$$

where

$$J = \overline{U_e \frac{\partial^2 \psi_t}{\partial x \partial y}} + \overline{\frac{\partial U_e}{\partial x} \frac{\partial \psi_t}{\partial y}} - \overline{\left( \frac{\partial \psi_0}{\partial x} + y \frac{\partial U_e}{\partial x} \right) \frac{\partial^2 \psi_t}{\partial y^2}} + \overline{\frac{\partial \psi_t}{\partial y} \frac{\partial^2 \psi_t}{\partial x \partial y}} - \overline{\frac{\partial \psi_t}{\partial x} \frac{\partial^2 \psi_t}{\partial y^2}} \qquad (40)$$

where the overbar denotes an average with respect to time.

Stuart (op. cit.) has assumed that the function J may be neglected giving

$$\frac{\partial \psi_{st}}{\partial y} \frac{\partial^2 \psi_{st}}{\partial x \partial y} - \frac{\partial \psi_{st}}{\partial x} \frac{\partial^2 \psi_{st}}{\partial y^2} = \nu \frac{\partial^3 \psi_{st}}{\partial y^3} \qquad (41)$$

This linearisation has allowed him to obtain a solution for the streaming layer. This solution is in qualitative agreement with the experiments of Schlichting (op. cit.) for a vibrating cylinder.

**The streaming flow**

A better model for the wall layer must be based on oscillating layers with a non-zero mean velocity. The analytical solution is very difficult and I have not found any successful attempt. Numerical solutions like the study of pulsatile blood flow by Schneck and Walburn (1976), the kernel studies mentioned previously or the DNS studies of the wall layer such as that of Schoppa and Hussain (2002) provide valuable information.

There are strong similarities between the results for laminar oscillating flow and turbulent flow DNS. Ejections of wall fluid imply boundary layer separation and this is only possible with the appearance of a zone of negative pressure behind the streaming jet as shown in the work of Schneck and Walburn. Johansson et al. (op. cit.) have found that the pressure patterns associated with shear layers near the wall undergo a development where an intense localised high-pressure region around and beneath the centre of the shear layer is found around the stage of maximum strength. At this stage, the maximum amplitude is about $2p_{rms}$ above the mean pressure. Johansson et al. suggest that these strong localised high-pressure regions could be of importance for boundary-layer noise generation. Strong shear layers are similarly produced on the upstream side of jets in cross flow where the pressure is high as in the forward stagnation region of a cylinder in a flow stream, (Chan, Lin, & Kennedy, 1976).

The ejections associated with bursting have been compared to jets of fluid essentially in cross flow to the main stream (Grass, 1971; A.A. Townsend, 1970; K.T. Trinh, 1992). The first difference to note is that unlike smoke plumes often studied as steady jets in cross flow, ejections from the wall layer are transient. The reason here is simple: the jets take fluid from the Stokes layer into the outer stream and therefore interrupt the source of the velocity fluctuations that feed the streaming flow.

Therefore the cause of the periodic inrush of fast fluid from the outer stream towards the wall is a consequence of the term $u_{i,0}u_{j,0}$ (Trinh, 1992) and not directly dependent of the term $\omega$. This non-oscillatory nature of the streaming flow is supported by the results of Schlichting (1960). Similarly Johansson et al. (op.cit.) found no signs of oscillatory motions or violent break-up in conjunction with shear layers embedded in the turbulent flow DNS database of Moin et al. (op.cit.) which, they believe, indicate a persistent motion of low-speed fluid away from the wall because they could follow the associated <U'V'> peaks for distances up to 1000 wall units.

The main crossflow deflects the wall ejection in a streamwise direction. This is the first interaction. The jet in crossflow has been divided into three zones as shown in Figure 7. The near-field region is jet dominated in the sense that the effects of the crossflow on the jet are not yet significant. In the curvilinear region, the initial jet momentum and the momentum extracted from the crossflow have comparable effects on the jet characteristics. In the far-field region, the effects of the crossflow predominate and the jet is aligned in the direction of the crossflow.

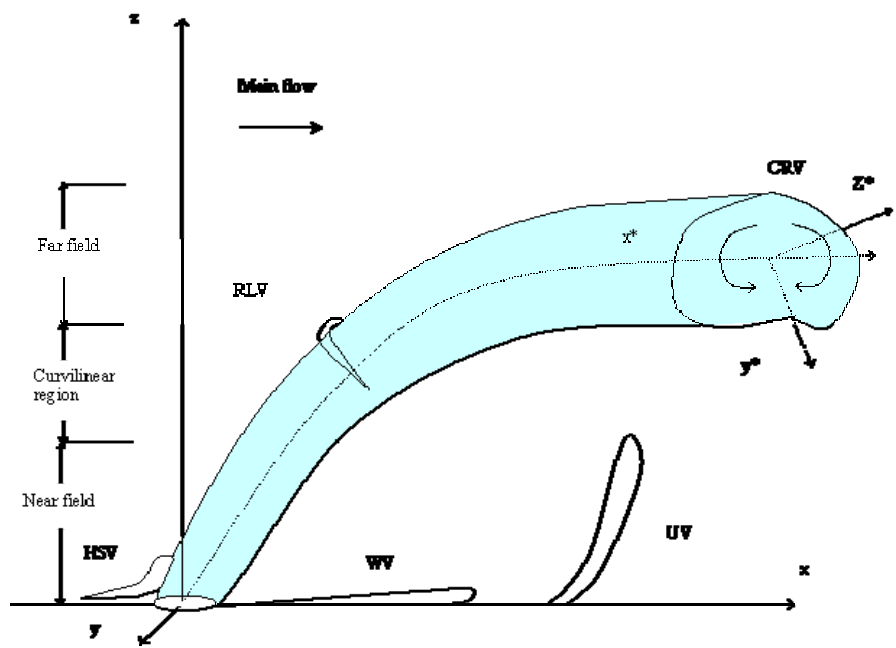

Figure 7. Geometry of a jet in cross flow. CRV Counter-rotating vortex, RLV Ring-like vortex, HSV Horseshoe vortex, WV Wall vortex, UV Upright vortex,, xyz

Cartesian coordinates, z*y*z* natural jet coordinates.

The second interaction between the mainstream and the jet is break up of the main flow. Johansson et al have also observed that the contribution of the Reynolds stresses to turbulence production in the downstream side of the shear layers is spatially spotty. This is compatible with the existence of a wake behind the ejections. In kernel studies mentioned previously, e.g. Peridier et al (op. cit.), the eruption of the laminar sub-boundary layer underneath the travelling vortex resembles the ejections and represents the intrusion of a stream of low-speed fluid into the outer inviscid region. Peridier et al have shown that a recirculation region exists behind the eruption. Liu, et al. (1991) have shown that the mainstream interacts with hairpin vortices near the wall and produce recirculation regions behind these hairpin vortices. There are other similarities between jets in cross flow and wall ejections. Falco (1977, 1991) has studied coherent structures in a boundary layer with smoke traces with the patterns shown in Figure 8. Falco observed two "typical eddy" forms: a mushroom shape on the back of the large coherent structures and a kidney shape near the edge of the boundary layer, which show striking resemblance to shapes observed with jets in cross-flow. The typical mushroom eddy is evident in the flow visualisation of plumes by Andreopoulos (1989) also reproduced in (Figure 8). The kidney shape represents a cross-section of the jet in the far field region where it is aligned in the direction of main flow. Townsend (1970) postulated that the ejections create roller like structures in the outer region. These have been deduced from probe measurements by Wark & Nagib (1991) who mapped out a recirculation zone associated with the roller-like structure behind the moving ejections (Figure 9) that is strikingly similar to the pattern obtained by Savory, Toy, McGuirk, & Sakellariou (1990) behind jets in crossflow.

The coherent structures created by a jet and the cross flow are in fact more complex and have received a large amount of attention in the last 30 years. Camussi, Guj, & Stella (2002) and Cortelezzi & Karagozian (2001) have summarised these structures, shown in Figure 7 as

1. CRVP (counter-rotating vortex pair) which is evident in the far field region
2. Ring-like vortices which are formed from the upwind shear layer of the jet flow

3. Horseshoe vortices formed upstream of the jet and close to the wall (very similar to the horseshoe vortices in the wall layer before the ejections)
4. WV, wall vortices which develop downstream of the jet orifice and close to the wall identified by McMahon, Hester, & Palfery (1971) and Fric & Roshko (1994)
5. UV, upright vortices that Fric and Roshko describe as "burst" of the boundary layer fluid.

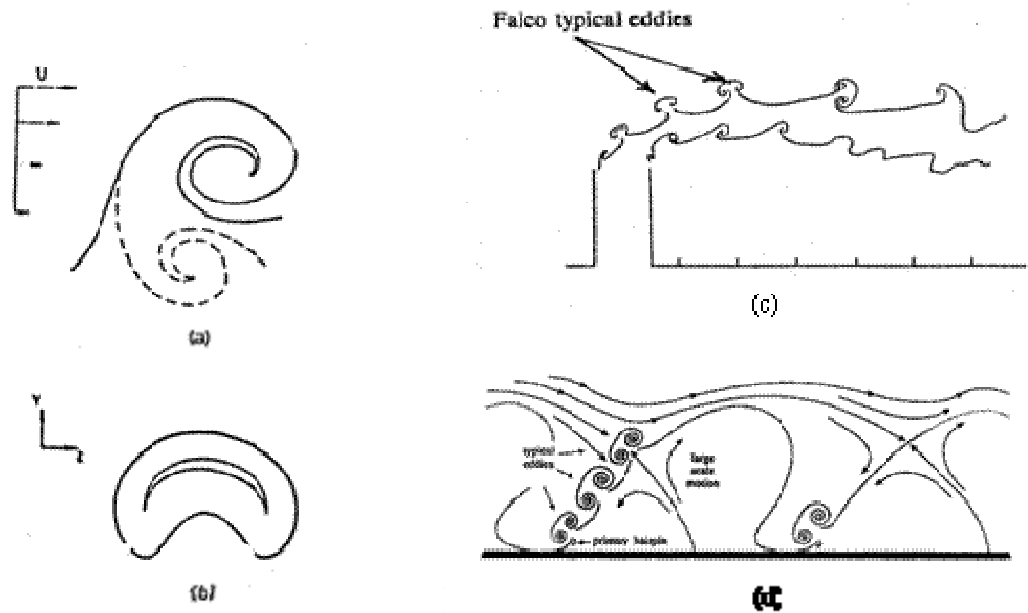

Figure 8 (a),(b) Typical eddies in turbulent boundary layers observed by Falco (1979), (c) Structure of plumes sketch of plumes after Andreopoulos (1989), (d) Sketch of ejection and typical eddies after Falco (1991)

**Influence of the interaction terms**

It is clear by now that in the study of turbulent flows we cannot ignore the effect of the interaction function J (equation (39) as Stuart (op. cit.) did in the study of oscillating laminar flow. If anything it should be further detailed. The writer submits that any study of turbulence that neglects this interaction effect will fail to reproduce the fine scale turbulence and require some sort of empirical closure model. So far only the DNS, which do not attempt to linearise the NS equations in that sense, have reproduced this interaction.

Nonetheless we can infer the character of these interactions without a formal analytical solution of the interaction function J. The effect of the interaction terms depends both on the inclination of the streaming jet that changes continuously with distance y from the wall, and the streamwise velocity that increases with distance y.

In the far field of the outer region when the jet path is aligned in the streamwise direction, the CRVP originates as an effect of the bending of the jet itself (Camussi et al., Cortelezzi et al. op.cit.) Because the jets in turbulent flow are intermittent, they represent unattached patches of fluids that would be more easily deflected in the streamwise direction. Presumably the CRVP of the ejections would have less interaction with the wake region that derives its vorticity mainly from the cross flow boundary layer, not the jet (Fric & Roshko, 1994). Smith and Mungal (1998) observed that jet fluid does not flow into the wake before a ratio of jet to cross flow velocities of 10.

The behaviour of the near wall region of jets in cross flow has been documented by many authors (Fric & Roshko, 1994; Kelso & Smits, 1995; Krothapalli, Lourenco, & Buchlin, 1990; McMahon, et al., 1971; Moussa, Trischka, & Eskinazi, 1977). Fric and Roshko argued that the near-wall flow around a transverse jet does not separate from the jet and shed vortices in the wake like the vortex shedding phenomenon from solid bluff bodies. They observed horseshoe vortices on the upstream side of the jet and argued that the vorticity in the wake region originates from the wall boundary layer flow which wraps around the jet and separates on its lee side creating wall vortices leading eventually to upright vortices that they describe as bursts. The horseshoe vortices are coupled with the periods of vortices that form in the jet wake (Krothapalli et al 1990, Fric and Roshko, 1994, Kelso and Smits 1995).

The horseshoe vortices upstream of the jet are strongly reminiscent of hairpin vortices widely observed in turbulent boundary layers and the wall vortices are similar to the longitudinal vortices in the sweep phase. I believe that the phenomena described here explain the key mechanism of self-sustenance of fully turbulent flows and can be captured with an adequate analysis of the interaction terms. I also agree with Schoppa and Hussain (2002) that the transition from laminar to turbulent flow does not

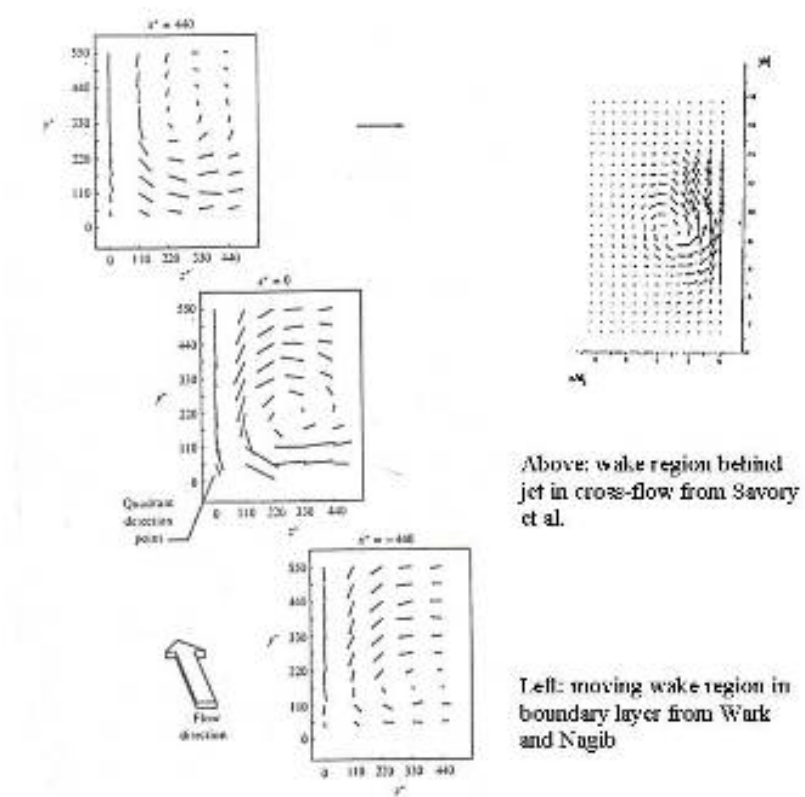

Above: wake region behind jet in cross-flow from Savory et al.

Left: moving wake region in boundary layer from Wark and Nagib necessarily require the original presence of a parent vortex and may be induced by travelling periodic waves. Nonetheless, once a streamwise horseshoe vortex has been formed, it will generate streaming flow and infant vortices downstream.

.

Figure 9. Roller-like structure in wake region behind a fixed jet in cross flow (Savory, et al., 1990) and moving with ejection (Wark and Nagib, 1991)

Of course there is a limit to the analogy between the jets in cross flow studied extensively in the literature and the ejections because the latter are intermittent. Thus the fluid at the wall in turbulent flows is not continuously supplied from upstream but must rush in from the log-law region to satisfy the equation of continuity as the ejections take the fluid in the low speed streaks away. Nonetheless, the streaming flow bursts are not instantaneous and there appears to be an overlap period when the adverse pressure created by the departing ejections can roll up the inrush fluid into streamwise vortices as observed by Kline et al. (1967).

## Energy flow in turbulence

In this visualisation, the flow of energy in a turbulent field may be described as follows. Energy is extracted from the main flow through periodic fluctuations, however they may be induced. The writer agrees with the arguments of Schoppa and Hussain (op.cit.) that the existence of a streamwise horseshoe vortex is not necessary to induce transition to turbulence. Even acoustic vibrations have been shown to induce streaming e.g. Schlichting p. 431(1960), Frater (1967). However, once self-sustenance of turbulent flow has been achieved, the main source of periodic fluctuations will be vortices travelling above the wall. The energy is stored in the sweep phase of the wall layer process in the form of growing wave fluctuations. This wave energy is then transformed into kinetic energy through the streaming process expressed in mathematical terms by the fast Reynolds stresses. When the magnitude of the fast Reynolds stresses reaches a critical threshold, fluid is ejected from the wall layer into the outer flow bringing with it the energy contained in the streaming flow. Fluid rushes into the wall layer after a burst to satisfy the law of conservation of mass. The energy contained in the streaming jets is dissipated eventually in the far field region through viscous interactions. Energy is also extracted from the cross flow to break it up when it impinges on the streaming jets. Some of that energy is dissipated as small scale turbulence but some is returned through the formation of "infant" vortices. In the wall layer, the infant vortices start a new sweep phase and perpetuate the process of turbulence generation.

## Advantages of a four component velocity decomposition

Examination of both theoretical analyses and experimental including DNS evidence show that even a basic solution requires us to express the instantaneous velocity in terms of at least 4 components:

$$u_i = U_i + \widetilde{U}'_i + u'_{i}(\omega t) + u_{i,st} \tag{42}$$

not just 3 as argued by Mankbadi (1992) and Schoppa and Hussain (2002). In fact there may be more components to add in the general case. For example, the fine scale turbulence observed in studies of jet in cross flow (Chan et al, op.cit.) may be captured by another velocity term that Schoppa and Hussain (op.cit.) call "incoherent" velocity in the sense that it does not lead to the formation of a new coherent structure. Even the simple analysis made here allows us to clearly identify the key features in the wall process and link them to particular terms in the NS.

**Subsets of the NS equations**

The wall layer flow before the advent of bursting obeys essentially the solution of order $\varepsilon^0$. It is well known in the study of oscillating boundary layers that the solution of order $\varepsilon^0$ is independent of the solution of order $\varepsilon$ and higher. Jimenez and Pinelli (1999) have similarly observed in their numerical experiment that "a cycle exists which is local to the near-wall region and does not depend on the outer flow". Stokes layers have been found embedded in many types of flow (Tetlionis, op. cit.). The oscillatory part of the solution of order $\varepsilon^0$ is described by the Stokes solution 2. Then when we set $\varepsilon = 0$ in equation (20) we obtain a subset of the NS equations that describes the penetration of negative viscous momentum from the wall into the main flow. This defines the thickness of the wall layer in turbulent flow. Essentially it describes the evolution of the smoothed velocity $\tilde{u}_i$ in the low-speed streaks. Analysis based on this simple subset of the NS equations reveals surprising insight into the classical statistical measures of turbulence (K. T. Trinh, 2009), p.48-80.

The growth of the fast fluctuations as the low-speed streaks develop can be analysed by the famous Orr-Somerfield equations (Orr, 1907; Sommerfield, 1908). A separate subset can be identified for the streaming flow and its interactions with the main flow.

The simplicity with which we can extract a picture of what happens is its main virtue but the solution itself is only accurate to order $\varepsilon$. If we decide to include the terms of order $\varepsilon^2$ in equation (24), the mathematical analysis becomes much more complex. Schlichting (1960) has discussed streaming flow in terms of the method of successive approximations in the study of non-steady boundary layers but does not even bother discussing third–and-higher approximations except to mention that the mathematical difficulties increase exponentially. Nevertheless one would suspect that this neglect of higher order terms immediately come at the cost of missing out on higher interactions between secondary and may be even tertiary and higher level structures.

We may also use the method of Reynolds (1895) to time-average the NS equations expressed in terms of the velocity expressed by equation (42). Clearly the resulting equations will include many more terms than the RANS. In particular there will be dedicated terms for the fast Reynolds stresses and their interactions with the main flow. These can be modelled specifically in computational fluid dynamic software giving more realistic profiles of turbulent flow fields.

There is wide consensus that the bursting (ejection) process is a defining characteristic of turbulent flows. If we drop the streaming term $u_{st}$ from equation (42) the solution can only describe oscillating laminar flow. If we further drop the fluctuating term the fast fluctuating term $u'_i$ the solution describes unsteady state laminar flow. If we further drop the term $\widetilde{U}'_i$ we obtain the subset of the NS equations for steady state laminar flow.

## A  Definition of Turbulence

To a beginner, the study of turbulence is immediately hampered by the surprising lack of a clear and concise definition of the physical process. Tsinober (2001) has published a long list of attempts at a definition by some of the most noted researchers in turbulence. The most common descriptions are vague: "a motion in which an irregular fluctuation (mixing, or eddying motion) is superimposed on the main stream" (Schlichting 1960), "a fluid motion of complex and irregular character" (Bayly, Orszag, & Herbert, 1988) or negative as in the breakdown of laminar flow in Reynolds' experiment (1883). Some of the definitions are quite controversial like Saffman's (1981) "One of the best definition of turbulence is that it is a field of random chaotic vorticity" because the words random and chaotic would imply that a formal mathematical solution, which is necessarily deterministic, does not exist. Perhaps the most accurate definition can be attributed to Bradshaw (1971) "The only short but satisfactory answer to the question "what is turbulence" is that it is the general-solution of the Navier-Stokes equation". This definition cannot be argued with but it is singularly unhelpful since no general solution of the NS yet exists 160 years after they were formulated.

The writer proposes that turbulence should be defined as "a system with a main cross flow containing secondary intermittent streaming, at some angle to the direction of the main flow and with which it interacts".

# Conclusion

The decomposition of the local instantaneous velocity in turbulent flows into four components is indicated by both theoretical and experimental considerations. The four components are a long-time average, a slow fluctuating component based on the difference between the long term average and the smoothed phase velocity developed by passing coherent structures, a fast fluctuating component which is periodic in nature and a streaming component created by the interaction between the fast fluctuations and the fluid viscous effects. This decomposition allows us to identify the terms in the NS equations that are associated with different structures and events in turbulent flows much more readily than the two-component decomposition of Reynolds.